# Redundancy in Systems which Entertain a Model of Themselves: Interaction Information and the Self-organization of Anticipation

Loet Leydesdorff
Amsterdam School of Communications Research (ASCoR), University of Amsterdam
Kloveniersburgwal 48, 1012 CX Amsterdam, The Netherlands.

**Abstract**
Mutual information among three or more dimensions ($\mu^* = -Q$) has been considered as interaction information. However, Krippendorff (2009a, 2009b) has shown that this measure cannot be interpreted as a unique property of the interactions and has proposed an alternative measure of interaction information based on iterative approximation of maximum entropies. $Q$ can then be considered as a measure of the difference between interaction information and redundancy generated in a model entertained by an observer. I argue that this provides us with a measure of the imprint of a second-order observing system—a model entertained by the system itself—on the underlying information processing. The second-order system communicates meaning hyper-incursively; an observation instantiates this meaning-processing within the information processing. The net results may add to or reduce the prevailing uncertainty. The model is tested empirically for the case where textual organization can be expected to contain intellectual organization in terms of distributions of title words, author names, and cited references.

## Introduction

This study follows up on a recent discussion about interaction information (Krippendorff, 2009a and 2009b; Leydesdorff, 2008a, 2009). In this discussion, Leydesdorff (2009, at p. 683) argued that the concurrent generation of probabilistic entropy (with "the arrow of time" (Coveney & Highfield, 1990)) and redundancy (against the arrow of time) in complex systems could be made measurable by considering the $Q$-measure as an imprint of meaning processing on information processing. If a modeling system, for example, feeds back from the perspective of hindsight on the modeled one by providing the latter with meaning, a complex interaction between information and redundancy generating mechanisms can be expected. Might it be possible to measure the outcome of this balance as a difference?

Krippendorff (2009b) responded to this question for another interpretation of $Q$ by elaborating the formula $R = I - Q$ (or equivalently $Q = I - R$). In this formula, $R$ indicates redundancy which, in Krippendorff's formulation, is generated by a (Boolean) observer who assumes independence among three (or more) probability distributions, whereas $I_{ABC \rightarrow AB:AC:BC}$ measures the (Shannon-type) interaction information generated in the system in addition to the lower-level (i.e., bilateral) interactions. In other words, $R$ is generated at the level of the model entertained by an observer because of the assumption that the probability distributions are independent. Given this assumption of independence, the algebraic derivation of $Q$ is valid because circular relations among the variables cannot occur under this condition and all probabilities add up to unity. However, $R$ measures the error caused by this assumption if the distributions are not independent.

In this contribution to the discussion, I shall argue that this possibility of using $I$ and $Q$ to measure $R$ enables us to address the question of how to measure the impact of the communication of meaning in a system which provides meaning to the events by entertaining a model. The Boolean observer who makes an error can be considered as the simplest case of a model maker. Using $Q$, the error in the expectations can be quantified (in terms of bits of information) in addition to interaction information $I_{ABC \rightarrow AB:AC:BC}$ in the modeled system. In a final section, I provide empirical examples of how one can measure, for example, the effects of intellectual organization as an order of expectations of textual organization in interactions among (three or more) attributes of documents.

## Dually layered systems entertaining a model of themselves

Anticipatory systems were defined by Rosen (1985) as systems which entertain a model of themselves. Since the modeling module is part of the system to be modeled, this reasoning would lead to an infinite regress. However, Dubois (1998) showed that this problem can be solved by introducing incursive and hyper-incursive equations into the computation of anticipatory systems. Unlike a recursive equation—which refers for the computation of its current state to a previous state ($t - 1$)—an incursive equation refers also to states in the present among its independent variables, and a hyper-incursive one refers to future states as co-constructors of present states.

Dubois' (1998) prime examples are the recursive, incursive, and hyper-incursive versions of the logistic equation. The hyper-incursive version of this equation can be formulated as follows:

$$x_t = a x_{t+1}(1 - x_{t+1}) \qquad (1)$$

and analytically rewritten into:[1]

$$x_{t+1} = \tfrac{1}{2} \pm \tfrac{1}{2}\sqrt{[1 - (4/a)\, x_t]} \qquad (2)$$

This equation has two real roots for $a \geq 4$. As is well known, the (recursive) logistic equation is defined only for $a < 4$; the solutions become increasingly chaotic as $a$ approaches four. In other words, $a = 4$ functions as a divide between the domain of the recursive version of the equation (along the arrow of time) and the hyper-incursive

[1] The following steps are included in the derivation (Dubois, 1998:9):

$$x_t = a x_{t+1}(1 - x_{t+1}) \qquad (1)$$

$$x_t = a x_{t+1} - a x_{t+1}^2$$

$$a x_{t+1}^2 - a x_{t+1} + x_t = 0$$

$$x_{t+1}^2 - x_{t+1} + x_t / a = 0$$

version (against the arrow of time). Furthermore, the incursive version of this equation generates a steady state for all values of *a* as follows: $x = (a-1)/a$.[2] This continuous curve crosses the noted divide between the two domains (Leydesdorff, 2008b; Leydesdorff & Franse, 2009).

In general, a model advances in time with reference to the present state of the modeled system. In other words, the model contains a prediction of a next state of the system. Hyper-incursive equations show how models may additionally feed back on the present state of the modeled system. The total system—that is, the system including its modeling subroutines—develops over time and thus generates entropy.[3] The model, however, reverses the time axis and can thus be expected to generate redundancy.

In other words, a model provides meaning to the modeled system. Meaning can be communicated in a reflexive discourse entertaining different models: a discursive model enables us to feed back on the system under study. This hyper-incursive meaning processing—formulated in Equation 1 exclusively in terms of possible future states—can be instantiated incursively by a reflexive observer who herself operates with an expectation in the present and historically with reference to a previous state. The model is thus instantiated (Giddens, 1984).

The communication of meaning cannot be expected to be directly measurable because, unlike the communication of information, hyper-incursive communication of meaning generates only redundancy. When the communication of meaning occurs within the system, however, this communication at a next-order level may make a difference for the communication of information by being instantiated locally *within* a historical process of information exchanges. The latter process necessarily generates probabilistic entropy, while the next-order level of meaning exchange can from this perspective also be considered as an exogenous network variable potentially restricting the options (Monge & Contractor, 2003). The model constrains the number of possible states in the system and can thus be considered as a filter reducing the uncertainty that prevails.

## The measurement of redundancy and interaction information

Might one be able to measure the reduction of uncertainty in an information processing system which entertains a model of itself? The generation of redundancy (= negative entropy) in information exchanges has been studied as a possible outcome of mutual

---

[2] The steady state can be found by solving $x_t = x_{t+1}$ as follows:

$$x = ax/(1+ax)$$
$$x(1+ax) = ax$$
$$ax^2 + (1-a)x = 0$$
$$x = 0 \lor x = (a-1)/a$$

[3] The second law of thermodynamics holds equally for probabilistic entropy, since $S = k_B H$ and $k_B$ is a constant (the Boltzmann constant). The development of *S* over time is a function of the development of *H*, and *vice versa.*

information among three or more dimensions. Ulanowicz (1986, at pp. 142f.), for example, suggested using this local reduction of uncertainty as an indicator of "ascendency."

McGill (1954) originally derived this measure as *A'*, but later adopters have denoted it with *Q*. The advantage of using the opposite sign as proposed by McGill (1954, at pp. 101 and 108) is that *Q* can be generalized for any dimensionality as $T(:\Gamma) = \sum_{S \subseteq \Gamma} Q(S)$, whereas mutual information can be expected to change signs with odd or even numbers of dimensions (Krippendorff, 2009b, at pp. 671f.). Note that in the case of three dimensions—on which we will focus below as the simplest case—*Q* is equal to the negative of mutual information, which I will denote as $\mu^*$ (following Yeung, 2008, at pp. 59f.; cf. McGill, 1954, at pp. 101 and 108). I return to the issue of the sign more extensively below.

Mutual information or transmission *T* between two dimensions *x* and *y* is defined as the difference between the sum of uncertainties in the two probability distributions minus their combined uncertainty, as follows:

$$T_{xy} = H_x + H_y - H_{xy} \qquad (3)$$

in which formula $H_x = -\Sigma_x p_x \log_2 p_x$ and $H_{xy} = -\Sigma_{xy} p_{xy} \log_2 p_{xy}$ (Shannon, 1948). When the distributions $\Sigma_x p_x$ and $\Sigma_y p_y$ are independent, $T_{xy} = 0$ and $H_{xy} = H_x + H_y$. In all other cases, $H_{xy} < H_x + H_y$, and therefore $T_{xy}$ is positive (Theil, 1972, at pp. 59f.). The uncertainty which prevails when two probability distributions are combined is reduced by the transmission or mutual information between these distributions.

Yeung (2008, at pp. 59f.; Abramson, 1963, at p. 129) specified the corresponding information measure in more than two dimension $\mu^*$ which can be formulated for three dimensions as follows:

$$\mu^*_{xyz} = H_x + H_y + H_z - H_{xy} - H_{xz} - H_{yz} + H_{xyz} = -Q \qquad (4)$$

Depending on how the different variations disturb and condition one another, the outcome of this measure can be positive, negative, or zero. In other words, the interactions among three sources of variance may reduce the uncertainty which prevails at the systems level.

Krippendorff (2009a and b) convincingly showed that multivariate *Q*-measures of interaction information cannot be considered as Shannon-type information. Watanabe (1960) had made the same argument, but this was forgotten in the blossoming literature using *Q*-measures for the measurement of "configurational information" (cf. Jakulin, 2005). Yeung (2008, at p. 59) used the deviant symbol $\mu^*$ to indicate that mutual information in three or more dimensions is *not* Shannon-type information. However, Krippendorff specified in detail why the reasoning among the proponents of this measure as an information statistics (about probability distributions) is mistaken: when more than two probability distributions are multiplied, loops can be expected to disturb what seems algebraically a perfect derivation. The probabilities may no longer add up to unity, and in that case a major assumption of probability theory ($\Sigma_i\, p_i = 1$) is violated.

Building on earlier work, Krippendorff (1982, 1986, 2009a) proposed another algorithm which approximates interaction information in three or more dimensions iteratively by using maximum entropies. This interaction information (denoted in the case of three dimensions as $I_{ABC\rightarrow AB:AC:BC}$) is Shannon-type information: it is the surplus information potentially generated in the three-way interactions which cannot be accounted for by the two-way interactions.

As noted, Krippendorff (2009b) provided an interpretation of $Q$ expressed as $Q = I - R$. In this formula, $R$ indicates redundancy which is generated at the level of the model entertained by an observer on the basis of the assumption that the probability distributions are independent and therefore multiplication is allowed. Given this assumption of independence, the algebraic derivation of $Q$ is valid because circular relations among the variables cannot occur and all probabilities add up to unity. However, when this assumption is not true, a Shannon-type information $I_{ABC\rightarrow AB:AC:BC}$ is necessarily generated. $Q$ can be positive or negative (or zero) depending on the difference between $R$ and $I$.

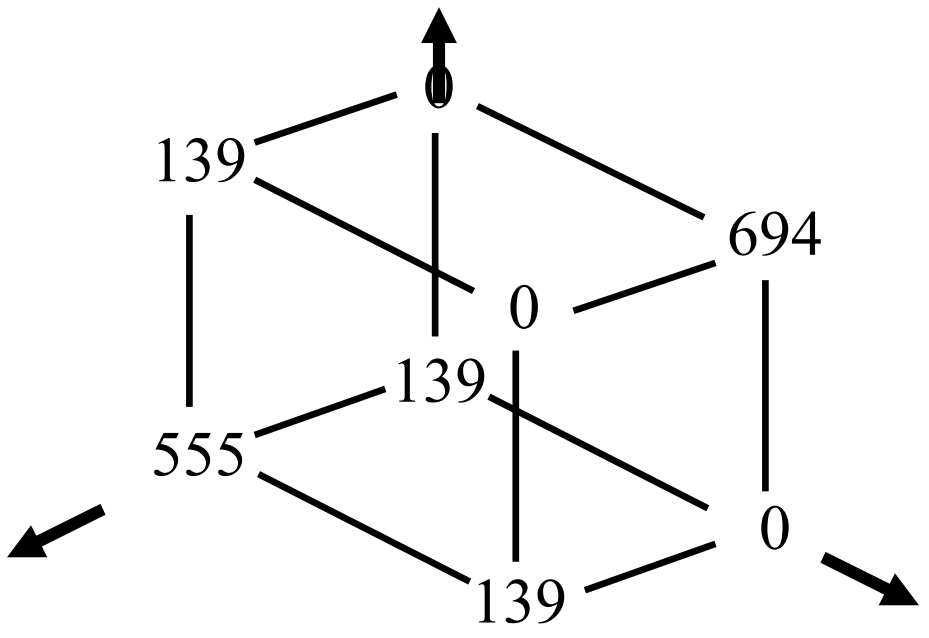

$Q(ABC)$ $= 0.00$
$I(ABC\rightarrow AB:AC:BC)$ $= 0.25$
$R(AB:AC:BC) = 0.25 - 0.00$ $= 0.25$

**Figure 1**: Example of positive ternary interaction with $Q = 0$ (Source: Krippendorff, 2009b, at p. 675).

In other words, $Q = 0$ should not be interpreted as the absence of interaction information, but only as an indicator of $R = I$. Figure 1 shows that the presence of (Shannon-type) interaction information $I$ is compatible with the absence of a value for $Q$. $Q$ cannot be interpreted as a unique property of the interactions. In other words, not $Q$, but $R$ is urgently to be provided with an interpretation. $Q$ can only be considered as the difference between the positive interaction information in the modeled system and the redundancy generated in the model.

## An empirical interpretation of *R*

*R* cannot be measured directly, but one can retrieve both *I* and *Q* from the data. For the measurement of $I_{ABC \rightarrow AB:AC:BC}$, Krippendorff's (1980, 1986, 2009a) iterative algorithm is available, and *Q* can be computed using Equation 4. Because $I_{ABC \rightarrow AB:AC:BC}$ is a Shannon-type information it cannot be negative. $\mu^*$ (= – *Q*) has to be added to (or subtracted from) *I* in order to find *R*.

Krippendorff (2009a, at p. 197) noted that in Shannon's information theory processes can flow only in a single direction, that is, with the arrow of time: "Accordingly, a message received could have no effect on the message sent." Meaning, however, is provided to the message from the perspective of hindsight, and meaning processing can be expected to operate orthogonally to—albeit potentially in interaction with—information processing. The modeling aims to reduce complexity and thus may add to redundancy without necessarily affecting the information processing. In the case of this independence, the model's expected information content remains "hot air" only specifying other possible states. However, if the meaning processing is made relevant to the information processing system, an imprint is generated. This imprint makes a difference and can be measured as *Q*, that is, the difference between interaction information in the modeled subsystem and redundancy in the modeling subroutine.

What type of (meta-)model might model this relation between information processing and meaning processing in numerical terms? Let me propose a rotated three-factor model as an example of a model of constructs that can be provided with meaning in terms of observable variables. The factor model assumes orthogonality among the (three or more) dimensions and one can measure the relations among these dimensions in terms of the factor loadings of the variables.[4] In other words, the data matrix represents a first-order network in which a second-order structure among the variables can be hypothesized using factor analysis.

The dimensions of the second-order structure are latent at the first-order level; they remain expectations about the main dimensions that span the network. By using orthogonal rotation among the factors, one can assure that the dimensions are analytically independent. Let us assume a three-factor model because the interactions among three dimensions provide us with the most parsimonious example for analyzing the problem of how redundancy because of data reduction operates in relation to interaction information.[5]

[4] The factor model is generated in terms of independent dimensions among the cases; this independence can be measured in terms of factor scores.
[5] Unlike principal component analysis—which essentially rewrites the variation in the matrix—factor analysis assumes the specification of a model.

The factor model spans a three-dimensional space in which the variables can be positioned as vectors. The variables are associated with the eigenvectors in terms of the factor loadings. (Factor loadings are by definition equal to Pearson correlation coefficients between variables and factors.) In other words, the (orthogonally rotated) factor matrix provides us with a representation of the variables in three main dimensions. Among these three dimensions one can compute an $I_{ABC \rightarrow AB:AC:BC}$, $\mu^*$, and therefore $R$. From this perspective, the difference between $R$ and $I$ can be considered as the remaining redundancy of the model that is not consumed by the Shannon information contained in the empirical distributions. If $R < I$, this difference can be considered as remaining uncertainty.

While $R$ is a property of the model and cannot be measured directly, the remaining redundancy or uncertainty ($R - I$) can be provided with *historical* meaning because it is manifest in the instantiation as $\mu^*$ (= $-Q$). Remaining redundancy can be considered as redundancy that could have been filled by the events. Remaining uncertainty would indicate that the model is not sufficiently complex to reduce all uncertainty in the data. Note that our meta-model assumes that the model generating $R$ is endogenous to the system as in the case of anticipation. The factor model can be considered as an example of a model that a system could entertain for understanding its own complexity.

One technical complication is that $\mu^*$ is measured as mutual *information* and not as redundancy. A negative value of $\mu^*$ in bits indicates a redundancy; a positive value of $\mu^*$ adds to the uncertainty. The inversion of the sign between $\mu^*$ and $Q$ may easily lead to confusion in empirical studies about what can be considered as reduction or increase in uncertainty. For example, Krippendorff (2009b, at p. 676) formulated: "With Q(ABC) = -1, redundancy measures R(AB:AC:BC) = 1 bit, which accounts for the redundant binary interaction in AB." This redundancy, however, would be equal to *minus* one bit when measured as information because adding to the redundancy reduces uncertainty at the systems level. In other words, if $I = 0$ then $R = Q$ because both $R$ and $Q$ are defined as redundancies. Hence, $R = I + Q$ or, more precisely, the value of $R$ (as a redundancy) = $I - \mu^*$ when the latter two terms are both measured in bits of information. When $\mu^*$ is measured as negative, this can be considered as an imprint—in this case, remaining redundancy—generated by a modeling system.

Since in our meta-model the model entertained by the system generating redundancy is also part of the information system, the "Boolean observer" (Krippendorff, 2009b, at p. 674) can no longer be considered as an external observer making an error. A second-order observer is part of the system as a recursive feedback loop (Maturana, 2000). Note that this second-order observer is embedded in the system, while entertaining our meta-model assumes an external (super-)observer (Maturana, 1978, at 56 ff.). The modeling within the system can be expected to involve a potential error when specifying an expectation because at the level of this model data are modeled and no longer assumed as given. The modeling subsystem can only specify an expectation about how the (three) orthogonal dimensions in the data interact historically, but this can be evaluated empirically—by an external observer—in terms of Shannon-type interaction information. This information is contained in empirical distributions to be observed (both internally and externally). The research question of this paper is whether one can show empirically how such reduction of uncertainty in the data would operate in a system which entertains a model of itself as a subroutine.

## Methods and materials

As data I use bibliometric data such as publications which can be related to one another in terms of citations, co-authorship relations, and co-occurrences of words in their titles. These various forms of relationships among documents enable us to construct matrices with the documents as cases (in the rows) and the different variables as columns. These matrices can be juxtaposed and combined. For example, in the first case—to be described below—the sample contains 102 documents (co-)authored by 183 authors containing 398 cited references, and 43 significant title words. We focus on the 24 authors who contributed more than once to the set.

| | $au_1$ | $au_2$ | … | $au_m$ |
|---|---|---|---|---|
| $doc_1$ | $a_{11}$ | $a_{21}$ | … | $a_{m1}$ |
| $doc_2$ | $a_{12}$ | … | … | $a_{m2}$ |
| $doc_3$ | … | … | … | … |
| … | … | … | … | … |
| … | … | … | … | … |
| $doc_n$ | $a_{1n}$ | … | … | $a_{mn}$ |

+

| | $w_1$ | $w_2$ | … | $w_k$ |
|---|---|---|---|---|
| $doc_1$ | $b_{11}$ | $b_{21}$ | … | $b_{k1}$ |
| $doc_2$ | $b_{12}$ | … | … | $b_{k2}$ |
| $doc_3$ | … | … | … | … |
| … | … | … | … | … |
| … | … | … | … | … |
| $doc_n$ | $b_{1n}$ | … | … | $b_{kn}$ |

=

| | $v_1$ | $v_2$ | … | $v_{(m+k)}$ |
|---|---|---|---|---|
| $doc_1$ | $c_{11}$ | $c_{21}$ | … | $c_{(m+k)1}$ |
| $doc_2$ | $c_{12}$ | … | … | $c_{(m+k)2}$ |
| $doc_3$ | … | … | … | … |
| … | … | … | … | … |
| … | … | … | … | … |
| $doc_n$ | $c_{1n}$ | … | … | $c_{(m+k)n}$ |

**Figure 2**: Two matrices for *n* documents with *m* authors and *k* words can be combined to a third matrix of *n* documents *versus* ($m + k$) variables.

Thus, we are able to construct three basic matrices (documents *versus* words, authors, or cited references, respectively) and various combinations (Figure 2 provides an example). This variation of matrices in substantively different dimensions enables us to specify different expected dimensions and accordingly to interpret the results of the (three-)factor model in substantive terms. In other words, using relational database management one is thus able to specify an expectation in terms of meaningful attributes.

The various data matrices are all subjected to the same procedures. First, we factor-analyze them in SPSS (v. 15) using Varimax rotation and choosing three factors for the extraction. Figure 3 provides an illustration of the three-factor matrix for the 48 title words mentioned above.

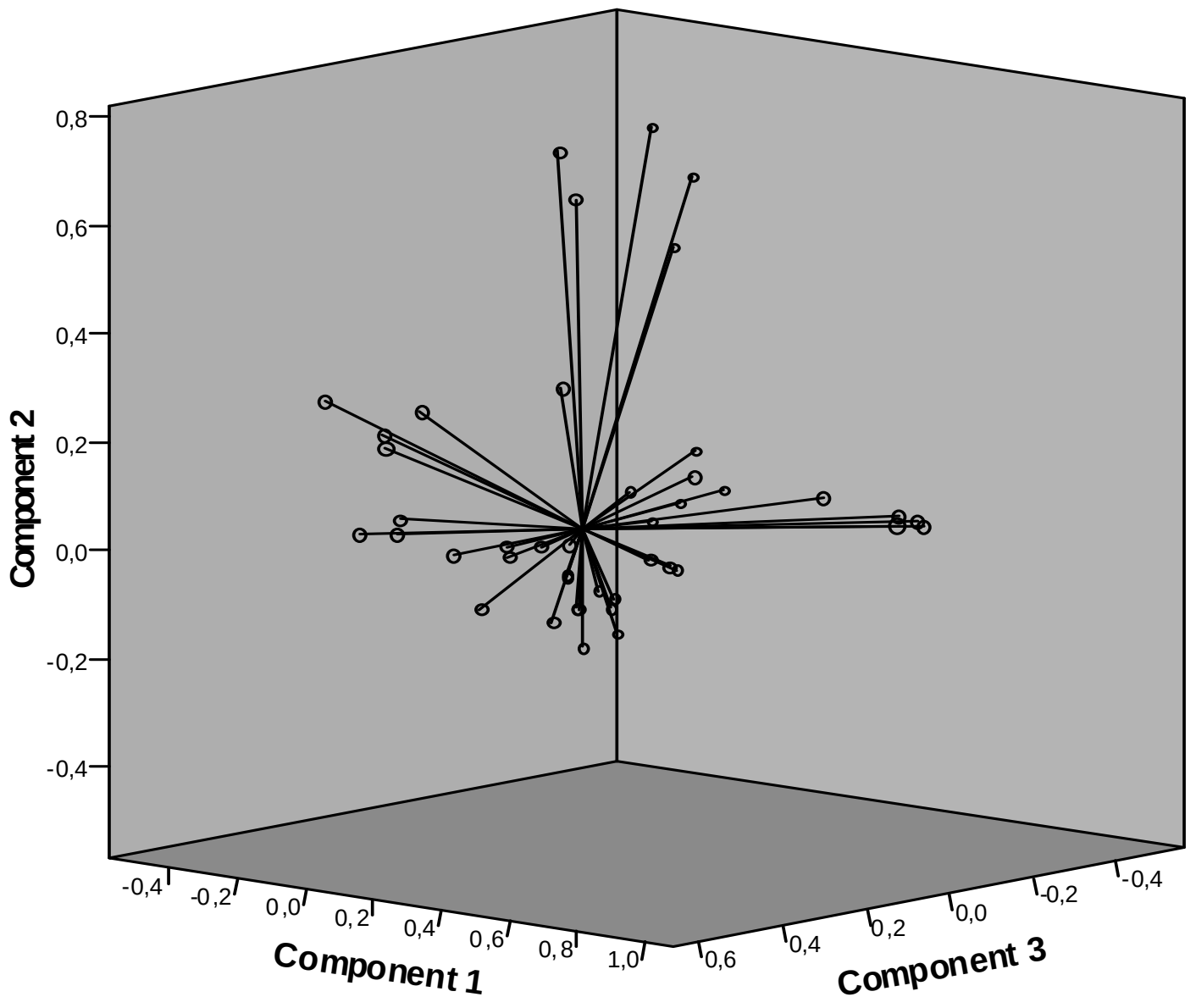

**Figure 3:** Forty-eight title words in rotated vector space. (Source: De Nooy & Leydesdorff, 2009.)

Each rotated component matrix contains three main (column) dimensions, and a number of variables (e.g., 48 in the above example) in the rows. The three dimensions can be considered as the orthogonal dimensions of a probability distribution. Using the rotated component matrix one can compute $\mu^*$ (or $Q$) and $I_{ABC \rightarrow AB:AC:BC}$, and therefore $R$ as the bias of the model. Note that the variables generate interactions by relating to the dimensions in terms of factor loadings.

In order to assess the data in terms of information theory, I subsequently bin the continuous data of the rotated factor matrix—with values between – 1.0 and + 1.0—in ten bins of 0.2. Because this is done in three dimensions, we operate on a probability distribution with $10^3$ cell values. This transformation into discrete data can be expected to generate another source of error. However, we develop our reasoning into empirical studies using bibliometric data. The various measures (*I, R,* and $\mu^*$) then can be appreciated in the bibliometrically informed analysis.

As data, I use two document sets with which I am familiar from other research efforts (De Nooy & Leydesdorff, 2009; Van Heurs *et al*., in preparation). The first dataset contains all 102 publications in the journal *Social Networks* in the period 2002-2008, downloaded at the *Web of Science* of the Institute of Scientific Information (ISI) on July 9, 2009. This set contains 398 cited references. Among the 183 (co-)authors we use the 24 authors who contributed more than once to the dataset, and we used the 43 (non-trivial) title words occurring more than twice.

The second dataset consists of 154 citable items (articles, reviews, letters, and proceedings papers) with publication years 2004 to 2008 in the journal *Social Studies of Science*. This data was downloaded on August 25, 2009. It contains 9,677 references, and the papers are authored by 201 (co-)authors of whom only ten contribute more than once. Of the 789 non-trivial title words, 72 occur more than twice and are included in the analysis.

**Results**

In both journals, the networks of authors and co-authors are composed of dyads and triads, and perhaps even larger groups of coauthors. Only recurrent authors can function as articulation points in the formation of a network. In the case of *Social Networks* there are 24 such authors; in the case of *Social Studies of Science* ten authors publish more than once.

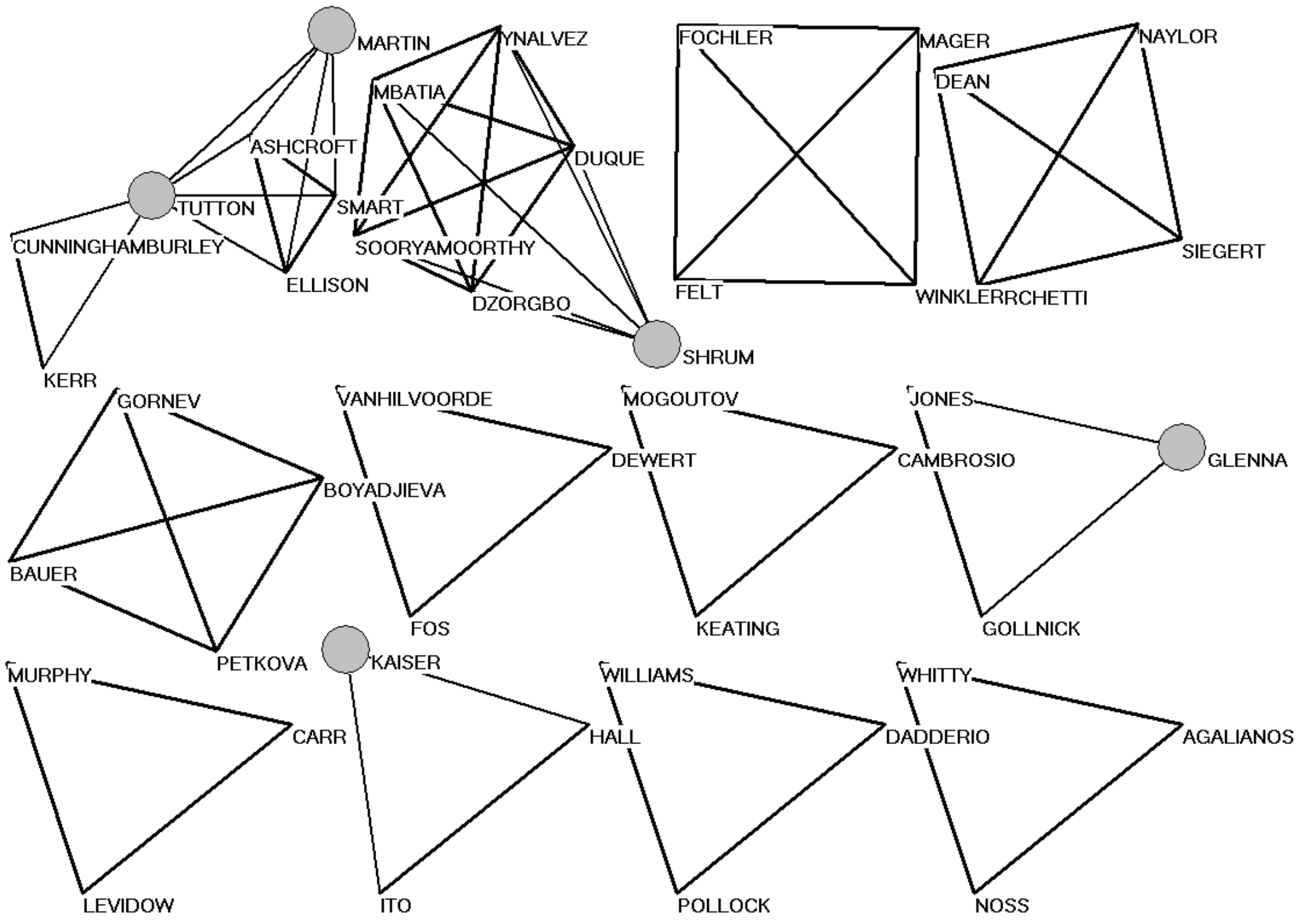

**Figure 4**: Triads and higher-order coauthorship patterns in *Social Studies of Science* (2004-2008).

Figure 4 shows the social network of triadic or higher-order co-authorship relations in the case of *Social Studies of Science.* (Dyadic relations and isolates are not depicted because they would overload the visualization.) The larger-sized nodes indicate five (among the ten) authors with more than a single publication in the set. The obvious lack of a large component in this social network (in terms of co-authorship relations in scientific publications) already indicates that networks of scholarly communication are organized not as social networks, but rather intellectually, for example, in terms of references and title words.

Title words in intellectually coherent sets can be used for the construction of semantic maps. Figure 5 provides such a cosine-normalized (Ahlgren *et al.*, 2003) map of the 43 title words used in the analysis of *Social Networks* (2006-2008).

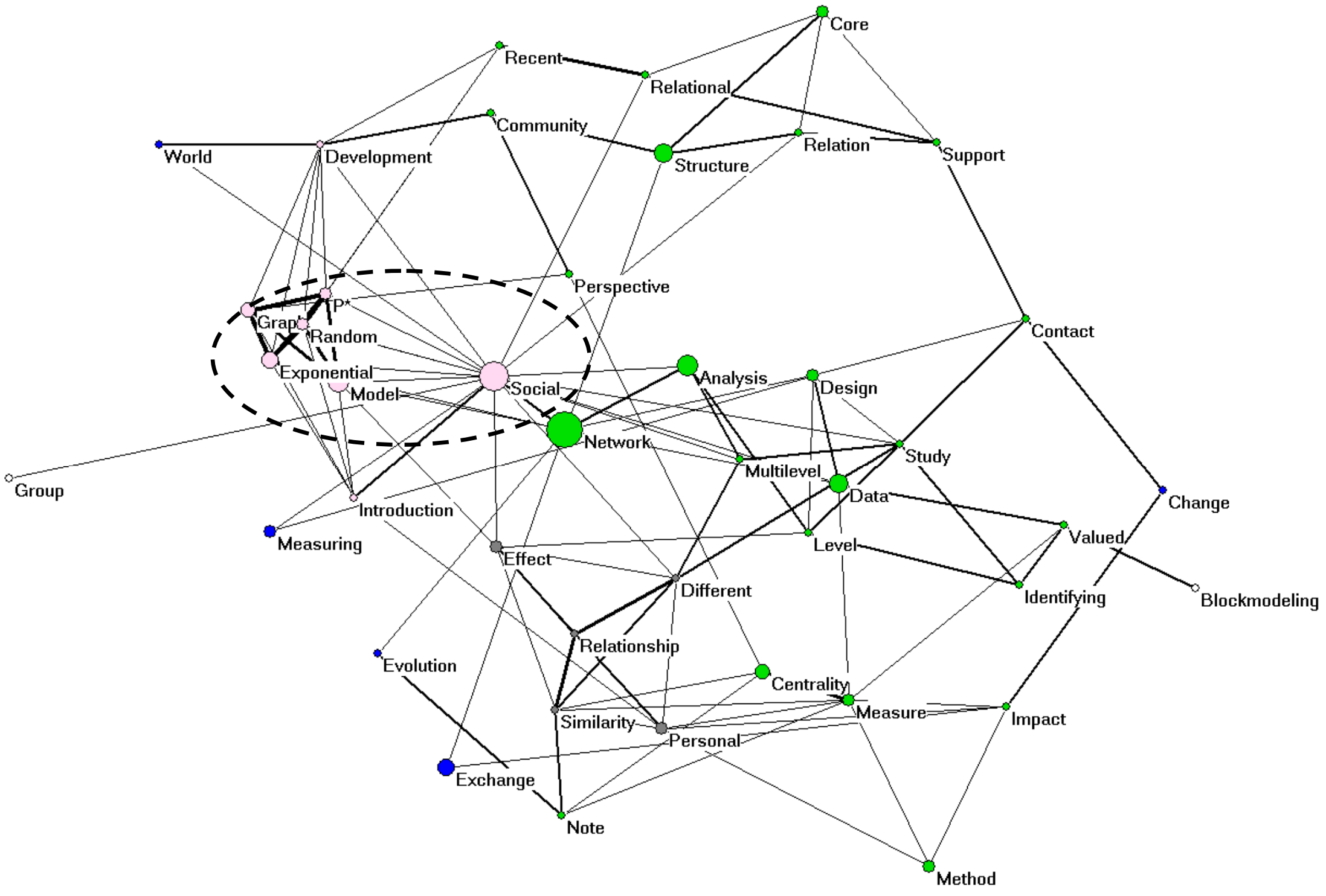

**Figure 5**: Cosine-normalized network among the 43 title words occurring more than twice in the document set of *Social Networks* (2006-2008); cosine ≥ 0.2. The size of the nodes is proportionate to the logarithm of the frequency of occurrence; the width of lines is proportionate to the cosine values; colors are based on the *k*-core algorithm; layout is based on energy minimization in a system of springs (Kamada & Kawai, 1989).

In this case, the use of the *k*-core algorithm in Pajek enables us to visualize, for example, a cluster of title words focusing on exponential random graph models.[6] This cluster is more related to the word "social" than to "network" given this algorithm. Note that the possibility of generating a meaningful semantic map upon visual inspection cannot yet validate the presence of analytical structure, since network visualizations are more flexible than analytical.

Cited references are more highly codified than title words in scholarly discourse or, in other words, words are more fluid than references (Leydesdorff, 1989). The journals and books used in references can form a tight network if the set is intellectually organized. The same set as in Figure 5, for example, can be used to draw Figure 6 on the basis of 395 of the 398 references connected to one another at the level of cosine ≥ 0.5.

---

[6] The network visualization program Pajek is freely available for academic usage at http://vlado.fmf.uni-lj.si/pub/networks/pajek/.

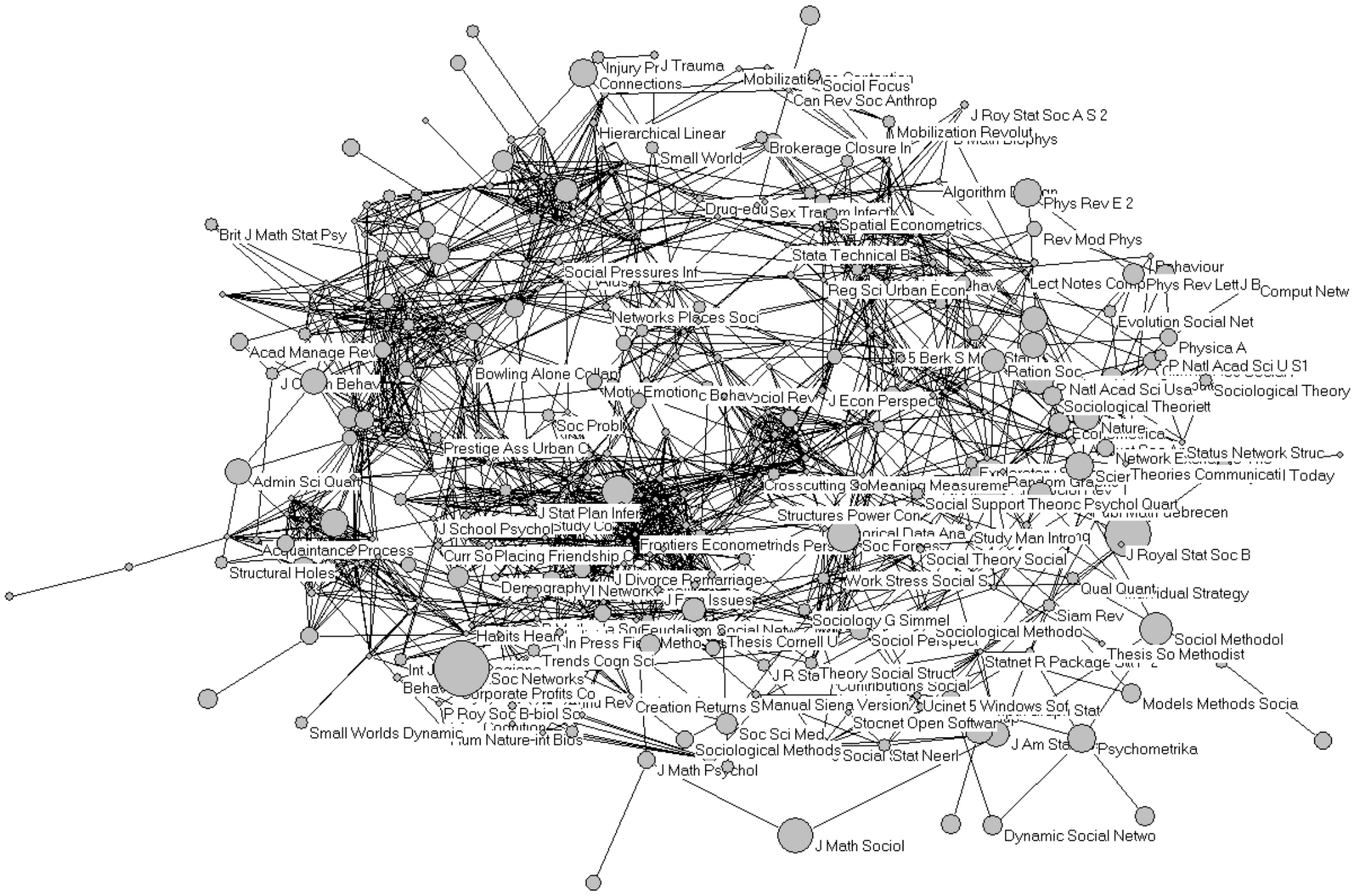

**Figure 6**: Map based on bibliographic coupling of 395 references in the 102 articles from *Social Networks*; cosine ≥ 0.5; Kamada & Kawai (1989). For the sake of readability a selection of 136 nodes (for the partitions 4 ≤ *k* ≤ 10) is indicated with legends.

In other words, maps based on cited references—this is also called bibliographic coupling (Kessler, 1963)—enable us to visualize the knowledge base in terms of which the authors of the documents in the set warrant their arguments. In this case, this set of references is tightly connected (cosine ≥ 0.5). In the case of *Social Studies of Science*, 9,677 references were included, indicating that in the qualitative and humanities-oriented tradition in sociology, references have a very different function: one can refer both to the texts under study and to studies relevant to a given approach (Leydesdorff & Akdag Salah, in preparation; Nederhof, 2006).

a. *Social Networks* (2006-2008)

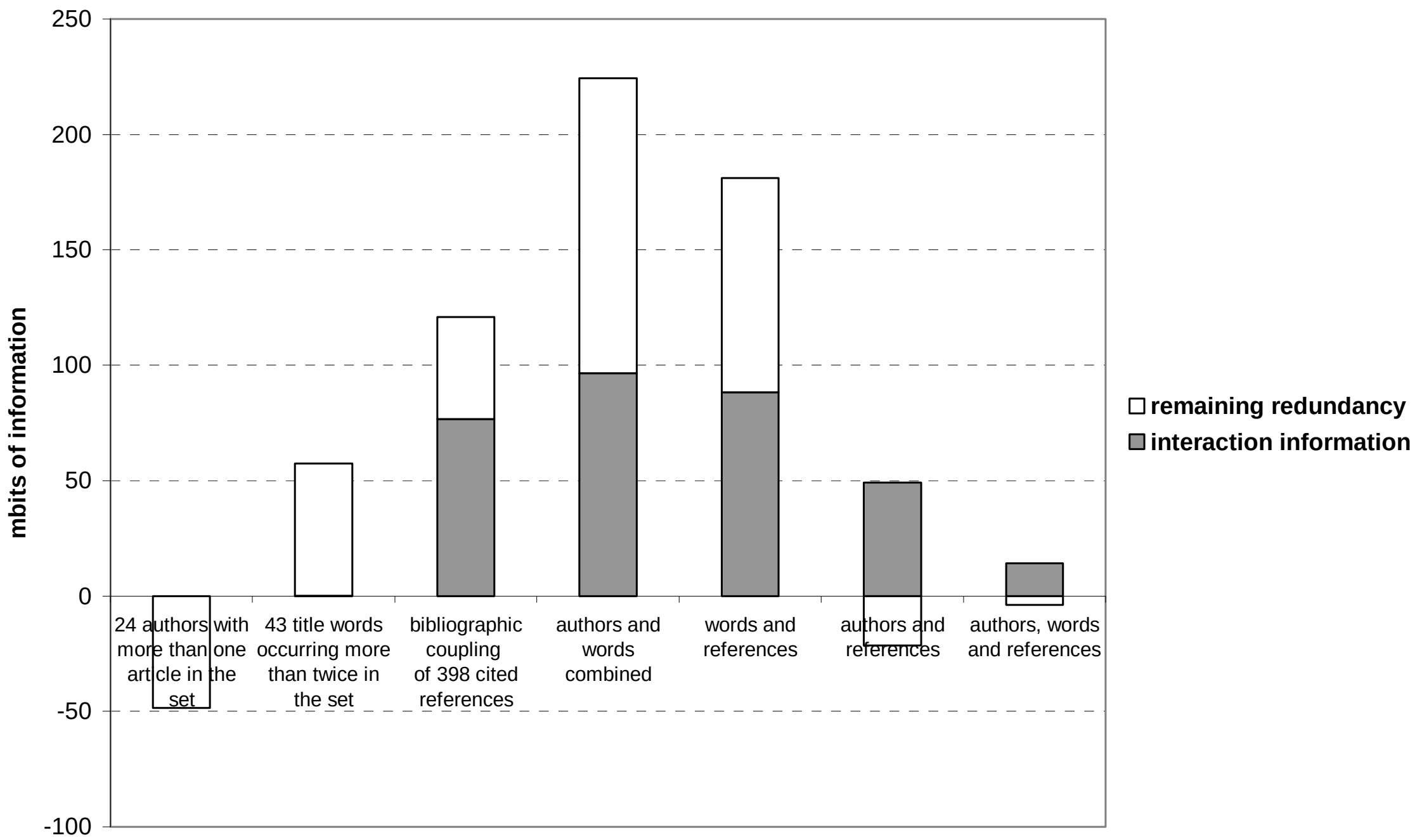

**Figure 7**: Interaction information ($I_{ABC \rightarrow AB:AC:BC}$) and remaining redundancy ($-\mu^*$ or $Q$) among the three main components in different dimensions and combinations of dimensions on the basis of *Social Networks* (2006-2008).

Figure 7 provides the results of the analysis of the (Shannon-type) interaction information among the three main components and remaining redundancy measured in terms of mutual information among these three dimensions. As can be expected on the basis of Figure 4, the network among the authors contains no second-order structure (the left-most bar in Figure 7). The redundancy generated by the factor model is negative in this case, and no interaction information is generated. The title words contain structure, as can be represented in a semantic map (Figure 7, second bar). However, this structure contains no interaction information; it remains historically volatile.

Bibliographic coupling in terms of titles of journals and books provides both structure in the model and interaction information. If one combines title words and author names as variables, both interaction information and remaining redundancy are considerably enhanced. This structure, however, is different from the one obtained by using the references as variables. Combining the three types of variables (words, authors, references; the right-most bar in Figure 7) does not lead to an improvement. The historical generation of interaction information remains (with the arrow of time), but the remaining redundancy becomes negative so that the assumed three-factor model can no longer account for the complexity observable in the empirical data.

b. *Social Studies of Science*

Figure 8 shows, analogously to Figure 7 and in accordance with the expectation, that the (co-)author network also in this case does not contain anticipatory structure and no interaction information among the main components is generated. If one focuses on the ten recurrent authors, the situation becomes even worse because these authors did not co-author among themselves.

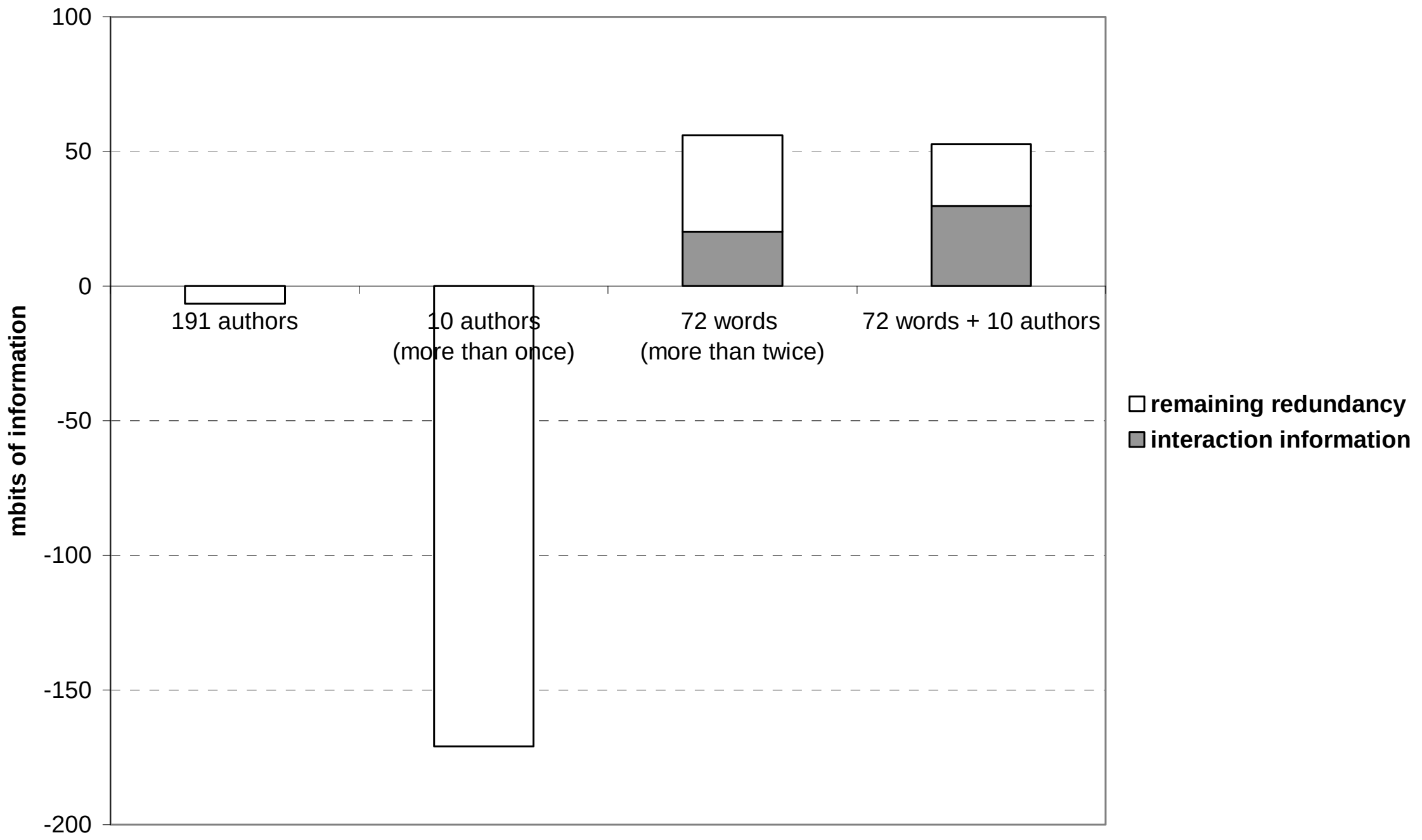

**Figure 8**: Interaction information ($I_{ABC \rightarrow AB:AC:BC}$) and remaining redundancy ($-\mu^*$ or $Q$) among the three main components in different dimensions and combinations of dimension on the basis of *Social Studies of Science* (2004-2008).

In this case of a qualitative and humanities-oriented journal, the words themselves can be considered as the carriers of codification. The structure among the meaningful title words generates both interaction information and redundancy. When the ten significant author names are added to this set, the interaction information in the historical basis is reinforced to the detriment of remaining redundancy. However, the total redundancy $R$ is not enlarged.

## Conclusions and discussion

I have argued that the two options of interaction information—McGill's (1954) algebraic and Krippendorff's (2009b) iterative approach—refer to two different contexts. The historical system containing three or more subdynamics can be expected while developing to generate all possible forms of Shannon-type information, including higher-order interaction information which cannot be reduced to lower-level (e.g., binary) interactions. This interaction information can be measured using Krippendorff's iterative algorithm specifying the maximum entropies of the distributions and their possible combinations. The algebraic approach, however, refers to a model that assumes independence among the probability distributions and therefore generates error when the distributions are not independent.

The model could first be considered as entertained by an external observer. The bias introduced by the model can be considered as expectations which can be compared with the measurement using the iterative approach. $Q$ (or $\mu^*$) measures the difference between the two. The situation changes conceptually if the observer is not external to the system, but contained in it as a second-order observer (Maturana & Varela, 1980; Von Foerster, 1982). The external observer can then be considered as a super-observer (Maturana, 178, at pp. 56 ff.), while the model entertained by the system reflexively follows the development of the system as an internal observer (Von Foerster, 1982). An observing system observes on the basis of entertaining a model. A system which entertains a model of itself can be considered as an anticipatory system (Dubois, 1998; Rosen, 1985), but the total system containing an observational model, can again be observed externally, for example, by an analyst as a super-observer.

The model is instantiated within the system by the observation at the locus of the embedded observer. Whereas this observer is embedded and therefore observes incursively, the model can be considered as analytical and operates hyper-incursively. It makes possible future states available to the system at the locus of observation. Since this feedback from a future state operates against the arrow of time, the second law is locally inverted. The modeled system, however, realizes I, and the difference can be measured as remaining redundancy (R – I) or remaining uncertainty (I – R). If I > R, the model is not able to process the complexity in the modeled system, and one cannot expect self-organization of an anticipatory layer to be preserved by the hypothesized dimensions. I illustrated this with examples of intellectual organization in textual domains.

In the empirical cases, the fruitfulness of this approach for understanding the relations between textual, social, and intellectual organization among scholarly communications was shown. In the case of a specialty with a focus on formalization, represented here by publications in the journal *Social Networks*, the relations among authors at the level of the social network tended to remain incidents and could not carry the intellectual system. The social relations are too uncertain for using a three-factor model. Words, however, can be used for the specification of a model as manifested in the semantic map.

Title words and their co-occurrences are not yet anchored in the historical development of the system—operationalized as *I*—whereas cited references are (Leydesdorff, 1989, 1997). The combination of title words and main authors, however, provided a codification which can operate historically, and, in this case, even more so than cited references. Furthermore, the cited references showed a structural dynamic different from the combinations of title words and (co-)author relationships: the two dynamics did not reinforce each other.

In the case of *Social Studies of Science*, a more humanities-oriented approach in the social sciences is archived. This first leads to an overwhelming number of references which cannot be expected to represent intellectual codification at the field level. In this case, title words and their co-occurrences could be shown to perform this function: both interaction information and remaining redundancy could be retrieved. The addition of the otherwise loose network among authors did not affect the redundancy *R* specified by the model, but the relative contribution of the realization of interaction information was larger. The author names help to anchor the intellectual organization which is manifested by title words and their co-occurrences, but this addition does not reinforce the structure already present in the semantic map.

In summary, I have wished to show how the intellectual organization of a specialty remains a model that feeds back as a possible set of realizations on the events that occur historically as textual organization or social networks among authors. The intellectual organization is not given, but constructed behind the backs of the constructors as an outcome of the interactions among them and their textual tools such as words and references. The model—the intellectual organization—remains in the sphere of anticipations or, in Husserl's (1929) terms, as *cogitata.* The realizations are historical and can be expected to generate Shannon entropy. The communication of meaning, however, can leave an imprint as a measurable redundancy or uncertainty depending on the sign of the difference.

## References

Abramson, N. (1963). *Information Theory and Coding*. New York, etc.: McGraw-Hill.

Ahlgren, P., Jarneving, B., & Rousseau, R. (2003). Requirement for a Cocitation Similarity Measure, with Special Reference to Pearson's Correlation Coefficient. *Journal of the American Society for Information Science and Technology,* 54(6), 550-560.

Coveney, P., & Highfield, R. (1990). *The Arrow of Time*. London: Allen.

De Nooy, W., & Leydesdorff, L. (2009). How can configurations be studied in 2-mode networks? Configurational information and "structuration". *Paper to be presented at the Two-Mode Social Network Analysis Conference, Free University Amsterdam, October 1-2, 2009*.

Dubois, D. M. (1998). Computing Anticipatory Systems with Incursion and Hyperincursion. In D. M. Dubois (Ed.), *Computing Anticipatory Systems, CASYS-First International Conference* (Vol. 437, pp. 3-29). Woodbury, NY: American Institute of Physics.

Giddens, A. (1984). *The Constitution of Society*. Cambridge: Polity Press.

Husserl, E. (1929). *Cartesianische Meditationen und Pariser Vorträge [Cartesian meditations and the Paris lectures]*. The Hague: Martinus Nijhoff, 1973.

Jakulin, A. (2005). *Machine learning based on attribute interactions* (Vol. http://stat.columbia.edu/~jakulin/Int/jakulin05phd.pdf). Ljubljana: University of Ljubljana.

Kamada, T., & Kawai, S. (1989). An algorithm for drawing general undirected graphs. *Information Processing Letters,* 31(1), 7-15.

Kessler, M. M. (1963). Bibliographic coupling between scientific papers. *American Documentation,* 14, 10-25.
Krippendorff, K. (1980). Q; an interpreation of the information theoretical Q-measures. In R. Trappl, G. J. Klir & F. Pichler (Eds.), *Progress in cybernetics and systems research* (Vol. VIII, pp. 63-67). New York: Hemisphere.
Krippendorff, K. (1986). *Information Theory. Structural Models for Qualitative Data.* Beverly Hills, etc.: Sage.
Krippendorff, K. (2009). Information of Interactions in Complex Systems. *International Journal of General Systems,* 38(6), 669-680.
Krippendorff, K. (2009). W. Ross Ashby's information theory: a bit of history, some solutions to problems, and what we face today. *International Journal of General Systems,* 38(2), 189-212.
Leydesdorff, L. (1989). Words and Co-Words as Indicators of Intellectual Organization. *Research Policy,* 18(4), 209-223.
Leydesdorff, L. (1997). Why Words and Co-Words Cannot Map the Development of the Sciences. *Journal of the American Society for Information Science,* 48(5), 418-427.
Leydesdorff, L. (2008a). Configurational Information as Potentially Negative Entropy: The Triple Helix Model. *Entropy,* 10(4), 391-410; available at http://www.mdpi.com/1099-4300/1010/1094/1391.
Leydesdorff, L. (2008b). The Communication of Meaning in Anticipatory Systems: A Simulation Study of the Dynamics of Intentionality in Social Interactions. In D. M. Dubois (Ed.), *Proceedings of the 8th Intern. Conf. on Computing Anticipatory Systems CASYS'07* (Vol. 1051 pp. 33-49). Melville, NY: American Institute of Physics Conference Proceedings.
Leydesdorff, L. (2009). Interaction Information: Linear and Nonlinear Interpretations. *International Journal of General Systems,* 38(6), 681-685.
Leydesdorff, L., & Franse, S. (2009). The Communication of Meaning in Social Systems. *Systems Research and Behavioral Science,* 26(1), 109-117.
Leydesdorff, L., & Akdag Salah, A. A. (in preparation). Maps on the basis of the Arts & Humanities Citation Index: the journals *Leonardo* and *Art Journal*, and "Digital Humanities" as a topic.
Maturana, H. R. (1978). Biology of language: the epistemology of reality. In G. A. Miller & E. Lenneberg (Eds.), *Psychology and Biology of Language and Thought. Essays in Honor of Eric Lenneberg* (pp. 27-63). New York: Academic Press.
Maturana, H. R. (2000). The Nature of the Laws of Nature. *Systems Research and Behavioural Science,* 17, 459-468.
Maturana, H. R., & Varela, F. (1980). *Autopoiesis and Cognition: The Realization of the Living*. Boston: Reidel.
McGill, W. J. (1954). Multivariate information transmission. *Psychometrika,* 19(2), 97-116.
Monge, P. R., & Contractor, N. S. (2003). *Theories of Communication Networks*. New York, etc.: Oxford University Press.
Nederhof, A. J. (2006). Bibliometric monitoring of research performance in the social sciences and the humanities: A review. *Scientometrics,* 66(1), 81-100.

Rosen, R. (1985). *Anticipatory Systems: Philosophical, mathematical and methodological foundations*. Oxford, etc.: Pergamon Press.

Shannon, C. E. (1948). A Mathematical Theory of Communication. *Bell System Technical Journal,* 27, 379-423 and 623-356.

Theil, H. (1972). *Statistical Decomposition Analysis*. Amsterdam/ London: North-Holland.

Ulanowicz, R. E. (1986). *Growth and Development: Ecosystems Phenomenology*. San Jose, etc.: toExcel.

Van Heurs, B., Leydesdorff, L., & Wyatt, S. (in preparation). Turning to Ontology in STS? Turning to STS through 'Ontology'

Von Foerster, H. (1982). *Observing Systems* (with an introduction of Francisco Varela ed.). Seaside, CA: Intersystems Publications.

Watanabe, S. (1960). Information theoretical analysis of multivariate correlation. *IBM Journal of research and development,* 4(1), 66-82.

Yeung, R. W. (2008). *Information Theory and Network Coding* (Vol. http://iest2.ie.cuhk.edu.hk/~whyeung/post/main2.pdf). New York, NY: Springer.